\RequirePackage{fix-cm}
\documentclass[smallextended]{svjour3_1}     
\smartqed  
\usepackage{amsmath}
\usepackage{graphicx}
\usepackage{rotating}
\usepackage{pdflscape}
 \usepackage{mathptmx}      
\usepackage{latexsym}
 \journalname{Experimental Astronomy}
\begin{document}

\title{A Method for Establishing a Station-Keeping, Stratospheric Platform for Astronomical Research} 


\titlerunning{Station-Keeping Stratospheric Platform}        

\author{Robert Fesen \and Yorke Brown
}

\authorrunning{Fesen \& Brown}

\institute{6127 Wilder Lab \\ Department of Physics \& Astronomy \\
                 Dartmouth College, Hanover, NH 03755 USA \\
              \email{robert.fesen@dartmouth.edu}, \email{yjb@yorkebrown.com}
}

\date{Received: date / Accepted: date}

\maketitle

\begin{abstract}

During certain times of the year at middle and low latitudes, winds in the upper
stratosphere move in nearly the opposite direction than the wind in the lower
stratosphere.  Here we present a method for maintaining a high-altitude balloon
platform in near station-keeping mode that utilizes this stratospheric wind
shear.  The proposed method places a balloon-borne science platform high in the
stratosphere connected by a lightweight, high-strength tether to a ``tug''
vehicle located in the lower or middle stratosphere.  Using aerodynamic control
surfaces, wind-induced aerodynamic forces on the tug can be manipulated to
counter the wind drag acting on the higher altitude science vehicle, thus
controlling the upper vehicle's geographic location.  We describe the general
framework of this station-keeping method, some important properties required
for the upper stratospheric science payload and lower tug platforms, and
compare this station-keeping approach with the capabilities of a high altitude
airship and conventional tethered aerostat approaches.  We conclude by
discussing the advantages of such a platform for a variety of missions with
emphasis on astrophysical research. 


\end{abstract}

\section{Introduction}
\label{intro}

It has long been realized that a high-altitude observing platform located in
the stratosphere and thus above a significant fraction of the Earth's
atmosphere could offer image quality competitive with space-based platforms.
This was the motivation behind the series of Stratoscope I and II balloon
flights that ran in the late 1950s, 1960s, and early 1970s flying 0.3 to 0.9 m
telescopes to an altitude of 24 km (80 kft) and obtaining 0.2 arcseconds
resolution images of the Sun, planets, and selected stars and galaxies
\cite{Bahng59,Schwarz73,Woolf64}.  Stratoscope images along with recent
atmospheric turbulence studies \cite{Avila97,Habib06,Hoegemann04,Wilson03} have
shown that near diffraction-limited image quality can be achieved at altitudes
at or above 20 km (65 kft) where the telescope is above $\approx$ 95\% or more
of the atmosphere.

Since the end of the Stratoscope missions, few high-altitude balloon flights
have carried optical and near-infrared astronomical telescopes and detectors.
NASA's highly successful multi-million cubic foot, high-altitude balloons flown
at altitudes of 30 to 40 km (100 -- 130 kft) have largely been limited to the
Arctic and Antarctic summers and have typically involved heliophysics, x-ray,
gamma-ray, particle astrophysics, and IR/sub-mm programs that are unaffected by
daylight observing conditions.  Only a few high altitude balloon flights, like
the recent heliophysics SUNRISE telescope \cite{Solanki10}, have been conducted
outside of the polar regions.

However, such high altitude, daylight balloon missions are generally not suitable for
a broad spectrum of general astronomical observing programs requiring dark
sky observing conditions. The few nighttime high-altitude astronomical balloon
flights that have occurred have been limited to relatively short duration times
of a week or less \cite{Donas87,Perotti80,Welsh83}.

\section{Airships}

Despite an ever increasing number of space missions, there has been renewed
interest in recent years for exploring the use of high-altitude balloon flights
for nighttime astronomical research. This has resulted in a number of papers
discussing possible lighter-than-air (LTA) vehicles and telescope arrangements
for optical and infrared observations from non-polar locations
\cite{Bely95,Ford02,Hibbitts13,Roberts13,von97,von00}.  A self-propelled,
high-altitude, long endurance (HALE) stratospheric airship capable of keeping
station over a desired geographic location would be a highly attractive
platform for a variety of astronomical and other science missions
\cite{Fesen06}.  

A solar-powered airship operating at altitudes near 20 km, where the
stratospheric winds are lightest could, in principle, remain aloft for days,
weeks, or even months thus serving as a general purpose astronomical
observatory for night observations covering a broad set of targets having a
wide range of declinations.  Besides avoiding so-called ``no-fly zones''
over some countries that restrict free-floating balloon flights over their
territory, a station-keeping airship could provide simple and
continuous line-of-sight telemetry allowing for high-bandwidth data
communication to a single ground station.

In basic terms, a stratospheric airship differs from a conventional
airship or blimp in terms of cruising altitude, balloon fabric, and
propulsion.  Blimps have thick and robust gas envelopes, are flown at
relatively low altitudes ($< 3000$ m), at low speeds ($<15$ m/s), and are
powered by conventional piston engines.  Their advantage over airplanes is
their ability to stay aloft and hover for long durations without refueling and
to do so at a relatively low cost of energy consumption (see the recent
historical review of airships by Liao and Pasternak \cite{Liao09}).

The possibility of relatively low construction and operations costs have made
airships attractive for a host of potential uses.  For example, the US
Department of Defense (DoD) has funded several high-altitude airship designs
and test programs over the last decade with the goal of developing a reliable
low cost stratospheric, long duration platform which could provide wide area
surveillance and communications capabilities with good air defense.  Recent DoD
projects include Southwest Research Institute's (SwRI) Sounder and HiSentinel
vehicles \cite{Smith00,Smith11} and Lockheed-Martin's High Altitude Airship
(HAA) and High Altitude Endurance-Demonstrator (HALE-D) airships.

Unfortunately, despite considerable effort and expense, no self-propelled
airship built by any manufacturer has flown at stratospheric altitudes for more
than one day.  The current record for a high altitude airship flight duration
may still be the High Platform II vehicle built by Raven Industries and flown
in the late 1960s at 20.4 km (67 kft) for a few hours \cite{SR03}.

A 2007 NASA study of a variety of LTA and heavier-than-air (HTA) unmanned HALE
vehicles found LTA vehicle concepts attractive in terms of performance but were
viewed as carrying a high technical risk \cite{Nickol07}.  This assessment was arrived
at, in part, due to the fact that the design and construction of a
high-altitude airship poses several major obstacles including large envelope
size, extremely lightweight and fragile balloon fabric for lifting gas
containment, energy storage and power systems, launch and recovery operations,
diurnal thermal management, and high-altitude propulsion motors and propellers
\cite{Davey08}.

A more recent 2012 assessment of US military airship efforts (GAO Report 13-81)
also gave an unfavorable outlook for the future development and deployment of
high altitude airships.  In reviewing various recent HALE airship efforts, the
report noted that many have been either terminated or have suffered
``significant technical challenges, such as overweight components, and
difficulties with integration of software development, which, in turn, have
driven up costs and delayed schedules.'' 

Despite such setbacks, strong interest in the development of a high-altitude,
long endurance airship persists.  Several commercial telecommunication
companies continue to pursue HALE airship development because such platforms
could provide communication and data services to consumers in rural or remote
areas \cite{Davey08,Djuknic97,Platt99,Relekar02,Tozer01} and would combine some
of the best features of satellite and fixed wireless services such as short
transmission delay times, low propagation loss, and relatively large service
areas \cite{Grace05}.  Airship programs such as the recently completed European
HAPCOS project (http://www.hapcos.org), the Japanese Stratospheric Airship
Platform Study \cite{Equchi98}, the Google Internet balloon project (``Project
Loon''), and Thales Alenia Space Consortium's ``StratosBus'' are among some of the
more recent efforts to use balloons for telecommunications purposes.

One of the most difficult problems in airship design is propulsion power.
While stratospheric wind speeds are lowest (5 -- 15 m/s) at altitudes around 20
km (65 kft), wind can vary significantly both daily and throughout the year,
exceeding 25 m/s at times and even higher in gusts.  At these speeds, wind force on a
conventional natural shape balloon is considerable, driving airship
designers toward aerodynamic balloon shapes with low form drag values and
propulsion systems involving large solar arrays or hydrogen fuel cells.

The form or shape drag force $F_{form}$ acting on a vehicle moving through a
fluid of density $\rho$ at speed $v$ is \[ F_{form} = \frac{1}{2} \rho v^2 A_f
C_D \;,\] where $A_f$ is the drag area (equal to the projected frontal area) of
the vehicle and $C_D$ is the coefficient of form drag corresponding to the
particular shape of the vehicle.  Similarly, the frictional drag force
$F_{friction}$ is \[ F_{friction} = \frac{1}{2} \rho v^2 A_w C_{SF} \;,\] where
the area $A_w$ is the ``wetted surface" and the coefficient $C_{SF}$ is the
skin frictional drag coefficient (which depends on the viscosity of the fluid).

To illustrate the wind induced drag forces on an airship,
we will consider the HiSentinel50 airship built by SwRI.  This vehicle was cylindrical
in shape with length $L = 54$ m and diameter $D = 12$ m. Its 
frontal area was $A_f = 115$ m$^2$,  its wetted area was
$A_w = 2500$ m$^2$ with drag coefficients estimated at $C_D = 0.022$ and
$C_{SF} = 0.0026$.  At an altitude of 65 kft the air density is $\rho = 0.091$
kg/m$^3$ meaning that for a wind speed of 10 m/s, its total drag force is
\begin{align} F_{drag} &= F_{form} + F_{friction} \nonumber \\ &=
\textnormal{12 N + 30 N = 42 N} \;. \nonumber \end{align} This force is the
thrust needed to oppose its wind-induced drag.

The power the airship needs to match this wind force and thereby enable it to keep station is 
\begin{align}
P &= F_{drag} \times v \nonumber \\
  &= \textnormal{42 N $\times$ 10 m/s = 420 W}\;.\nonumber
\end{align}

This amount of power is relatively small and practical for an airship using
photovoltaic (PV) panels and lightweight electric motors. But this example
represents a fairly favorable scenario in terms of mild stratospheric winds of
just 10 m/s at the ``sweet spot" altitude around 20 km plus a very low drag
airship design.  Since drag is proportional to the square of velocity and power
is drag times velocity, propulsion power is really proportional to $v^3$.  Thus
airship power requirements increase rapidly with wind speed.  

For instance, using the same airship numbers above but now for a wind speed of
30 m/s, the airship's total drag force increases to nearly 370 N requiring
11 kW of power to keep station. This is a considerable amount of power to
generate in order to maintain the airship floating above its desired position
point, apart from any power that might be required by the airship's
payload. 

However, even at lower wind speeds, having an airship keep station could be
challenging.  If the airship's overall drag forces were twice as large due
perhaps to a larger form drag coefficient for the airship or caused by a large
and highly non-streamlined mission payload shape, the power required would
increase by a factor of two.  This would mean some 20 kW of power would then be
needed for station-keeping, an amount difficult to generate using PV panels
alone on this relatively modest sized airship. Since steady wind speeds around
30 m/s are not exceptionally rare at 20 km, this means that strict year-round
station-keeping for such an airship might simply not be possible. 

\section{Tethered High-Altitude Airships and Balloons}

A radically different approach for establishing a lighter-than-air
stratospheric station-keeping platform involves tethering the vehicle to a
ground station.  This scheme again would keep the platform's altitude to 20 km
or so as to take advantage of the lightest stratospheric winds and hence the
lowest drag forces on the airship.

However, no tethered high altitude stratospheric aerostat has been successfully
flown for even one full diurnal cycle, although several attempts were made by
French atmospheric scientists in the late 1970s \cite{Regipa74}.  The main
obstacles include aviation restrictions, tether strength and weight, the tether
winch, and tether wind drag.  Storms and wind gusts in the troposphere can
generate large transient wind loads on the tether, the winch, and the vehicle
itself especially during initial deployment and recovery.

Despite this, a tethered stratospheric aerostat offers some distinct advantages
over powered airships. These include no propulsion motors or propellers
allowing for higher mass payloads, no large solar panel arrays to power the
propulsion motors, and no large batteries for nighttime propulsion.  In
addition, the advent of technically advanced, high tensile strength materials
such as Ultra High Molecular Weight Polyethylene (UHMWPE) such as Spectra and
Dyneema), Polybenzobisoxazole (PBO) such as Zylon, and Liquid Crystal Polymers
such as Vectran, Kevlar, and Technora has made the concept of a tethered
stratospheric aerostat more practical than in the past. 
 
Several papers concerning the feasibility and flight properties of a tethered
aerostat at altitudes around 20 km have appeared recently. These include a
study of a sea-anchored stratospheric, long duration balloon
\cite{Akita12}, the construction, launch and operation of tethered
stratospheric balloons as alternatives for satellites \cite{Badesha02,Izet11},
and investigations of the dynamic response for a high altitude tethered balloon
aerostat and tether line to winds and their effects on payload pointing stability
\cite{Aglietti09,Grant96}. 
 
The chief advantage of the tethered LTA platform scheme is simplicity.  In
principle, a land or sea deployed tether to a stratospheric balloon from a
launch site with favorable tropospheric winds, few aviation hazards or flight
restrictions, and seasonal periods of low stratospheric wind speeds, might
allow flight durations exceeding a few days.  However, weather conditions
throughout the tropospheric column (e.g., surface and low altitude winds and
gusts, storms and downdrafts) along with tether mass and tether wind loading
may severely restrict its applicability and flight duration.

\begin{figure*}
\includegraphics[width=1.0\textwidth]{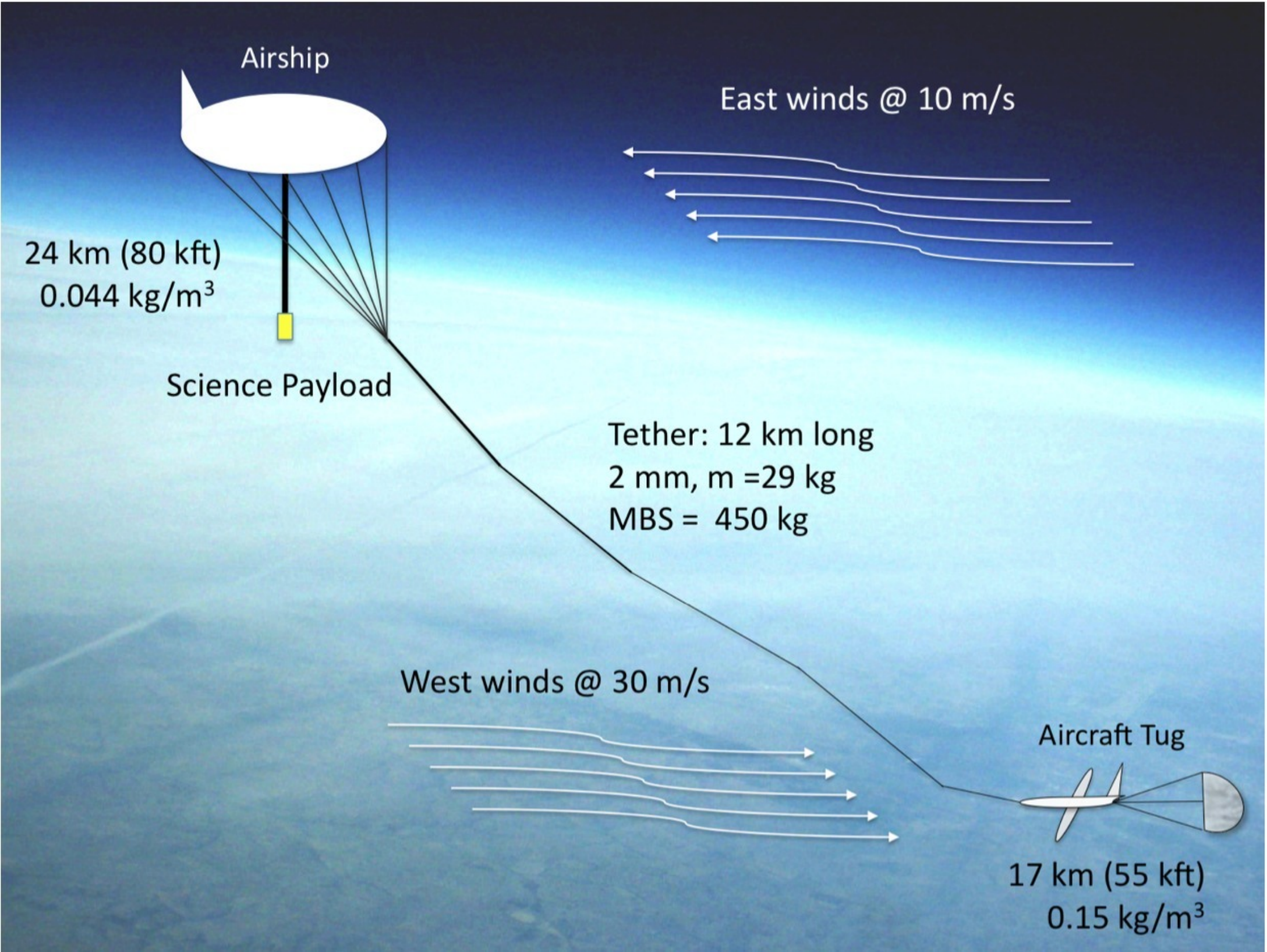}
\caption{Sketch of the concept for a stratospheric, station-keeping LTA science platform 
using naturally occurring East--West wind shear in the lower stratosphere.  An airship
carrying a science payload located at an altitude of 24 km (80 kft)
would experience generally lower wind speeds and in the opposite direction
than a ``tug" vehicle located in the  at 17 km (55 kft).
A lightweight tether would connect the two vehicles.
In the case shown here, a 2 mm diameter HMPE (Dyneema) cord is employed
with a minimum breaking strength (MBS) of 400 kg.
Station-keeping of the science platform would be accomplished by
active control of the aerodynamic forces acting on the tug.
In addition, some horizontal aerodynamic force on the science platform 
can be effected by a rudder or other control surface. }
\label{fig:1}       
\end{figure*}

As is done for low altitude aerostats, most high-altitude tethered airship
models have the tether attached to a ground-based winch which must be operated so as to 
limit the tension on the tether below its minimum breaking strength. Despite a
number of articles discussing this approach
\cite{Aglietti09,Akita12,Badesha02,Bely95,Izet11}, the only partially
successful series of flights seems to have been done by atmospheric researchers
in the 1970s \cite{Regipa74} and, to our knowledge, no high-altitude tethered
aerostats have been attempted since. 

\section{A Tethered Stratospheric Wind Shear Approach}

Here we describe an alternative means of establishing a stratospheric station-keeping
LTA platform that makes use of a tether.  During certain times of the year at
mid- and low latitudes, winds in the upper stratosphere move in nearly the
opposite direction than the wind in the lower stratosphere.  A balloon or
airship at high altitude could be tethered to a heavier-than-air glider ``tug"
at a lower altitude where the wind blows essentially in the opposite
direction.  By adjusting the aerodynamic configuration of the tug, wind
forces acting on it can be made to counteract those acting on the airship.

\begin{figure}
\includegraphics[width=1.0\linewidth]{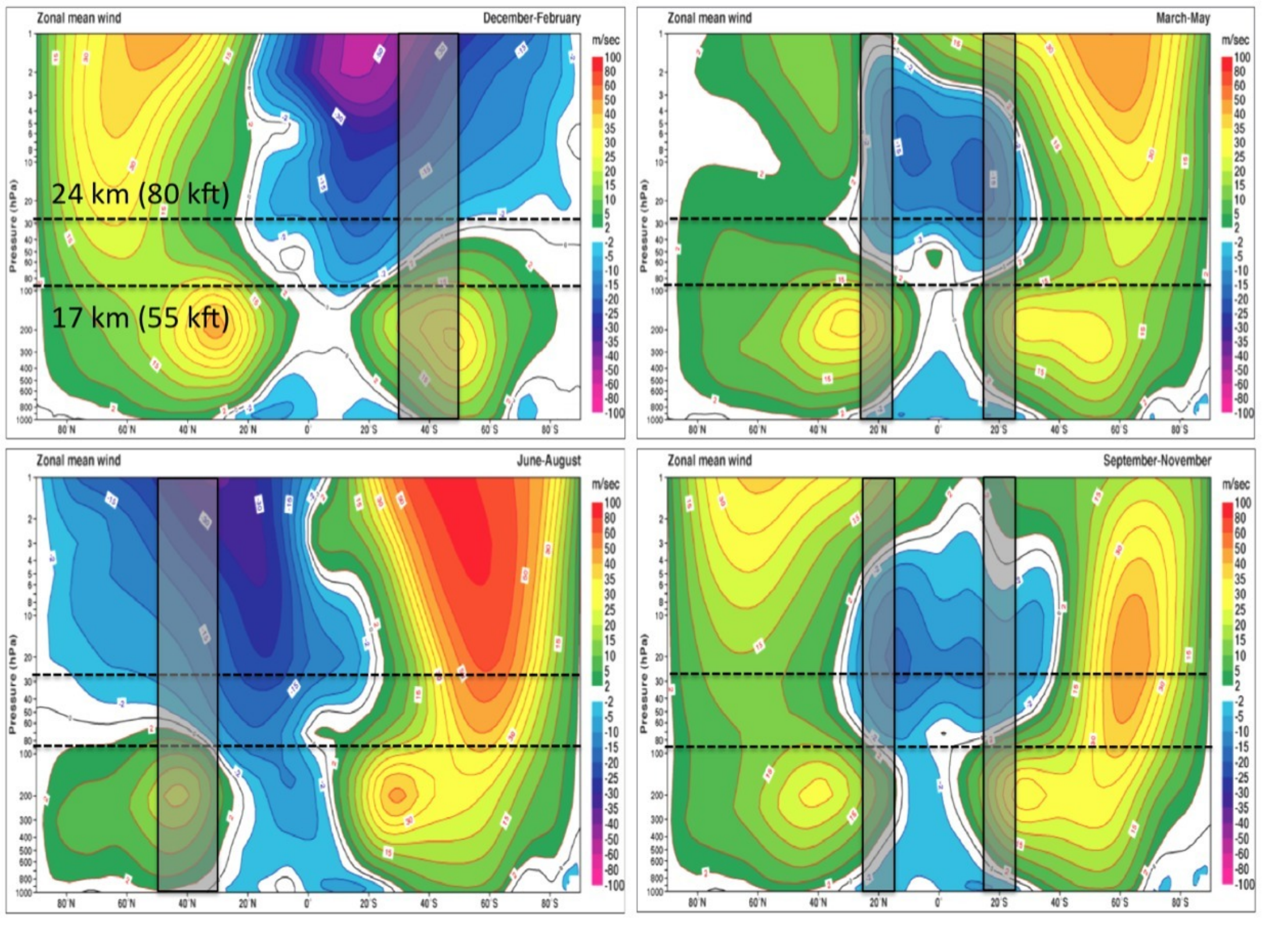}
\caption{Three month seasonal averages of tropospheric and stratospheric winds.
Each plot shows contours of wind speed (m/s) as a function of latitude
and pressure stratum (hPa).  The altitudes indicated correspond to
the example configuration discussed in the text and in Figure 1.
The shaded regions show ranges of 
latitudes where the wind shear is favorable for the example system.
In the northern hemisphere, optimum latitude varies by season from 15$^{\circ}$N in
winter to nearly 40$^{\circ}$N in summer.  The seasons and latitudes are reversed
in the southern hemisphere.  Adapted from plots taken from the
European Centre for Medium-Range Weather Forecasts (ECMWF) ERA-40 website. }
\label{fig:2}
\end{figure}

An example configuration exploiting this naturally occurring wind shear is
shown in Figure \ref{fig:1}.  The airship and its payload float at an altitude
around 24 km (80 kft) while the tug flies some 7 km lower at around 17 km (55 kft).
The tether connecting them is shorter and hence lighter than it would need to
be if it were to extend to the ground and it does not penetrate the turbulent
weather of the troposphere.  The tug's relatively high altitude places it well
above the maximum operating ceilings of all commercial aircraft (43 kft) and
private or corporate jets (51 kft) thereby greatly reducing aviation
restrictions and hazards.  Wind at the tug's altitude is generally stronger and
the air denser than higher up meaning the tug can be relatively small and still
develop the necessary forces to balance that experienced by the upper
airship.

This approach to a station-keeping capability depends upon 
stratospheric wind shear---that is, the difference in wind speed and direction
between the altitude of the airship and that of the tug.  Figure \ref{fig:2}
shows plots of wind speed and direction as a function of altitude and latitude
where altitude is indicated by the associated atmospheric  pressure.  Although
these plots are multi-year averages for each season, they illustrate the basic
stratospheric wind shear phenomenon.  Each plot is annotated with the example
altitudes discussed above and with a range of latitudes for which favorable
conditions prevail.

\begin{figure}
\includegraphics[width=0.52\linewidth]{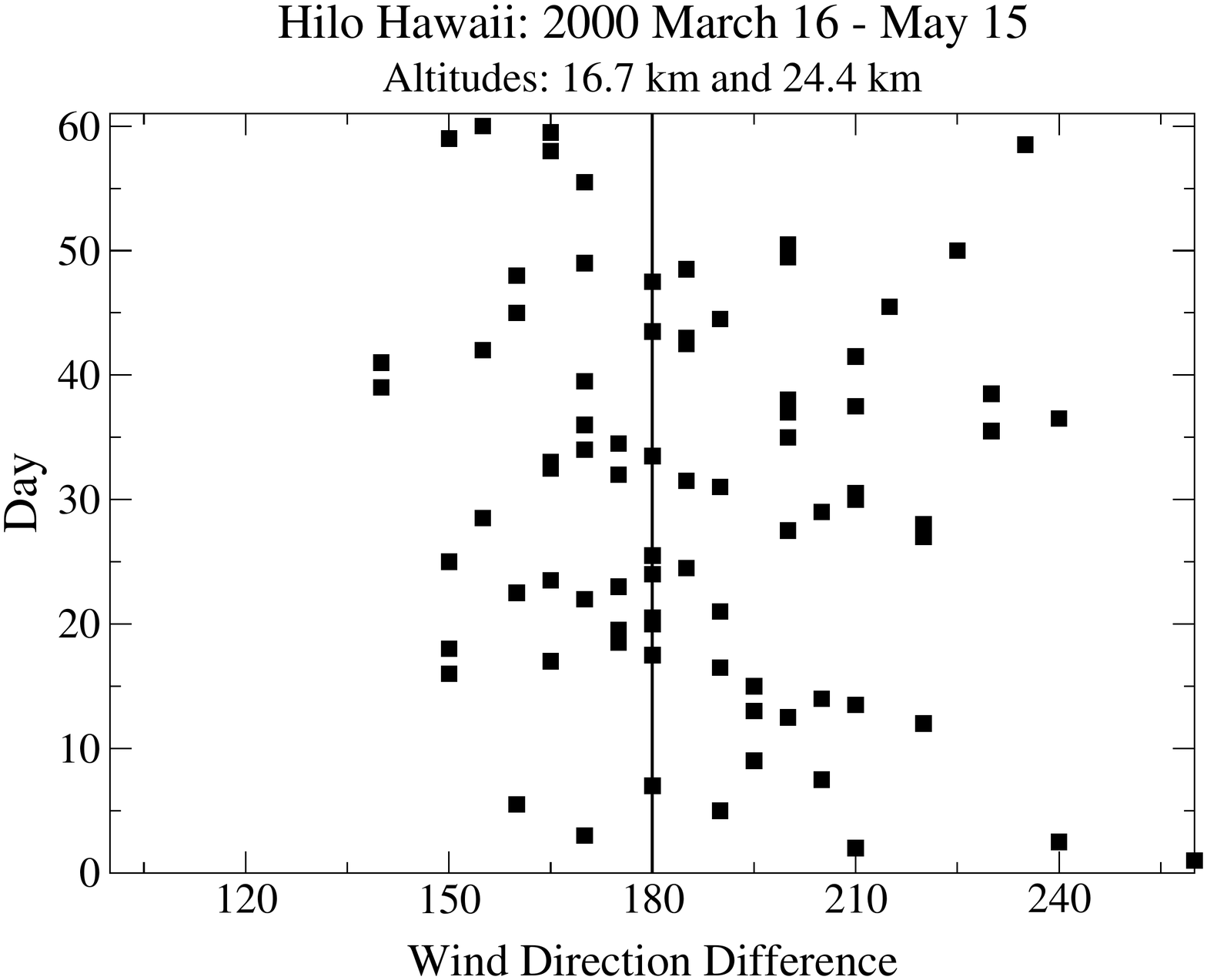}
\includegraphics[width=0.52\linewidth]{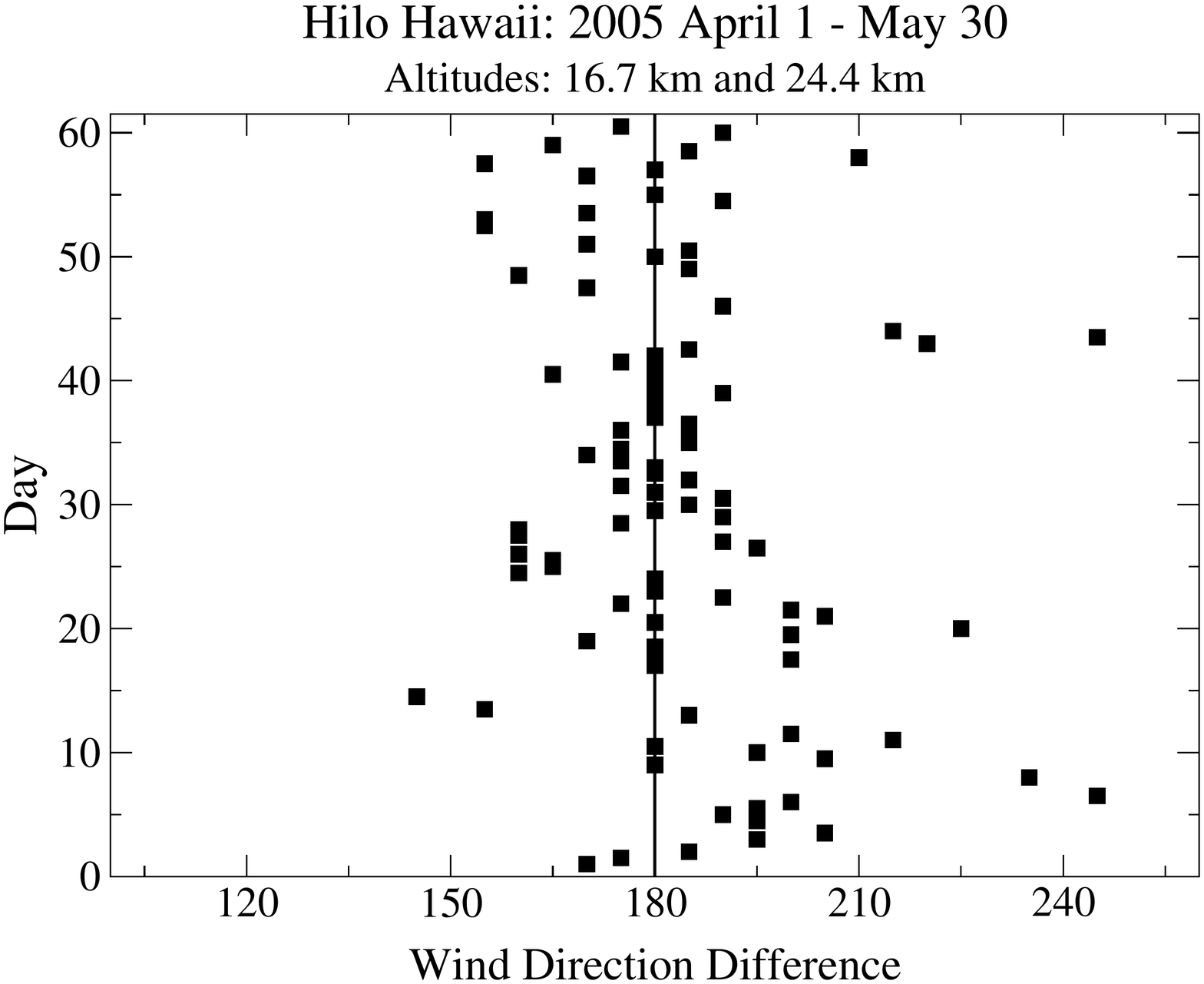} \\
\includegraphics[width=0.52\linewidth]{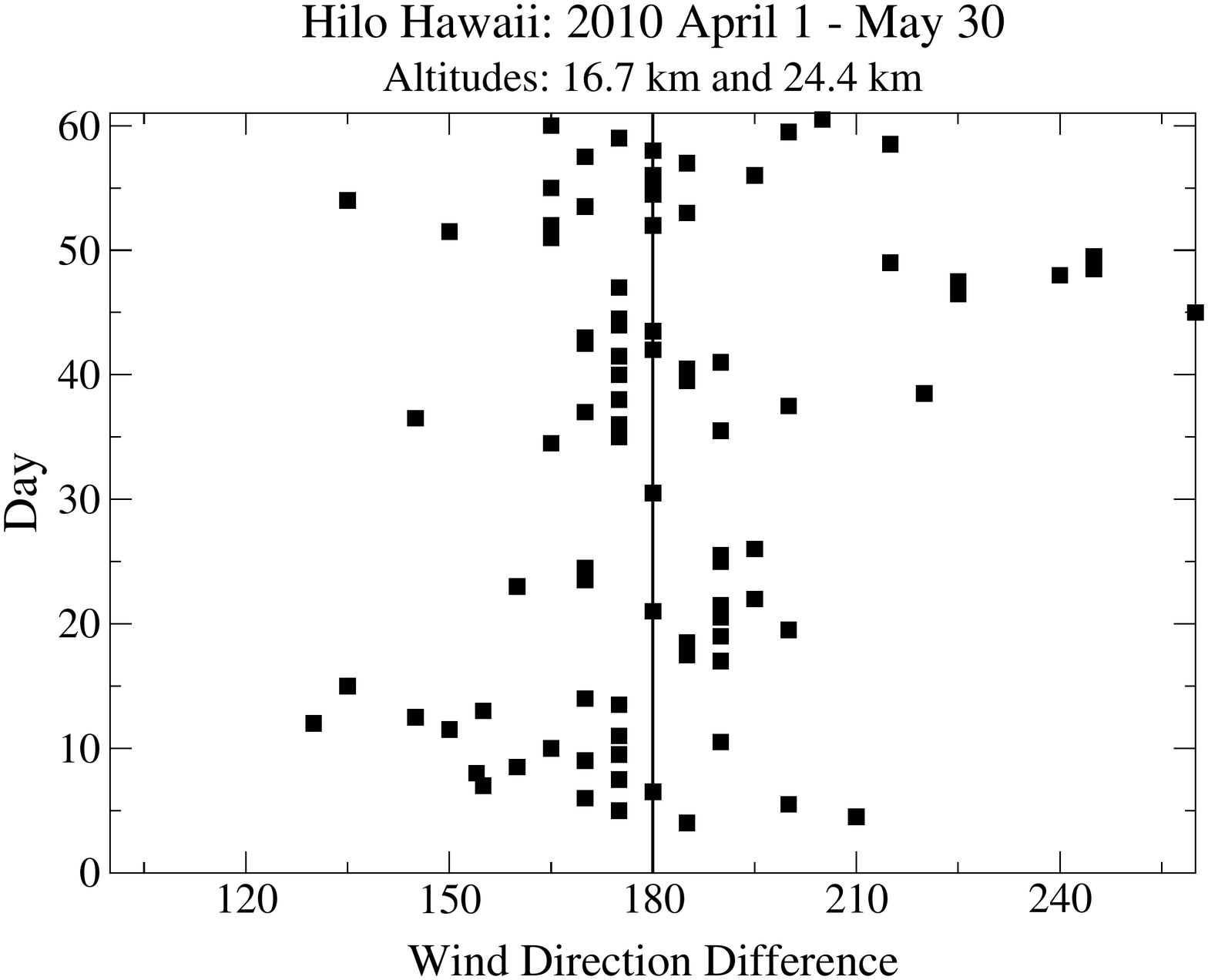}
\includegraphics[width=0.52\linewidth]{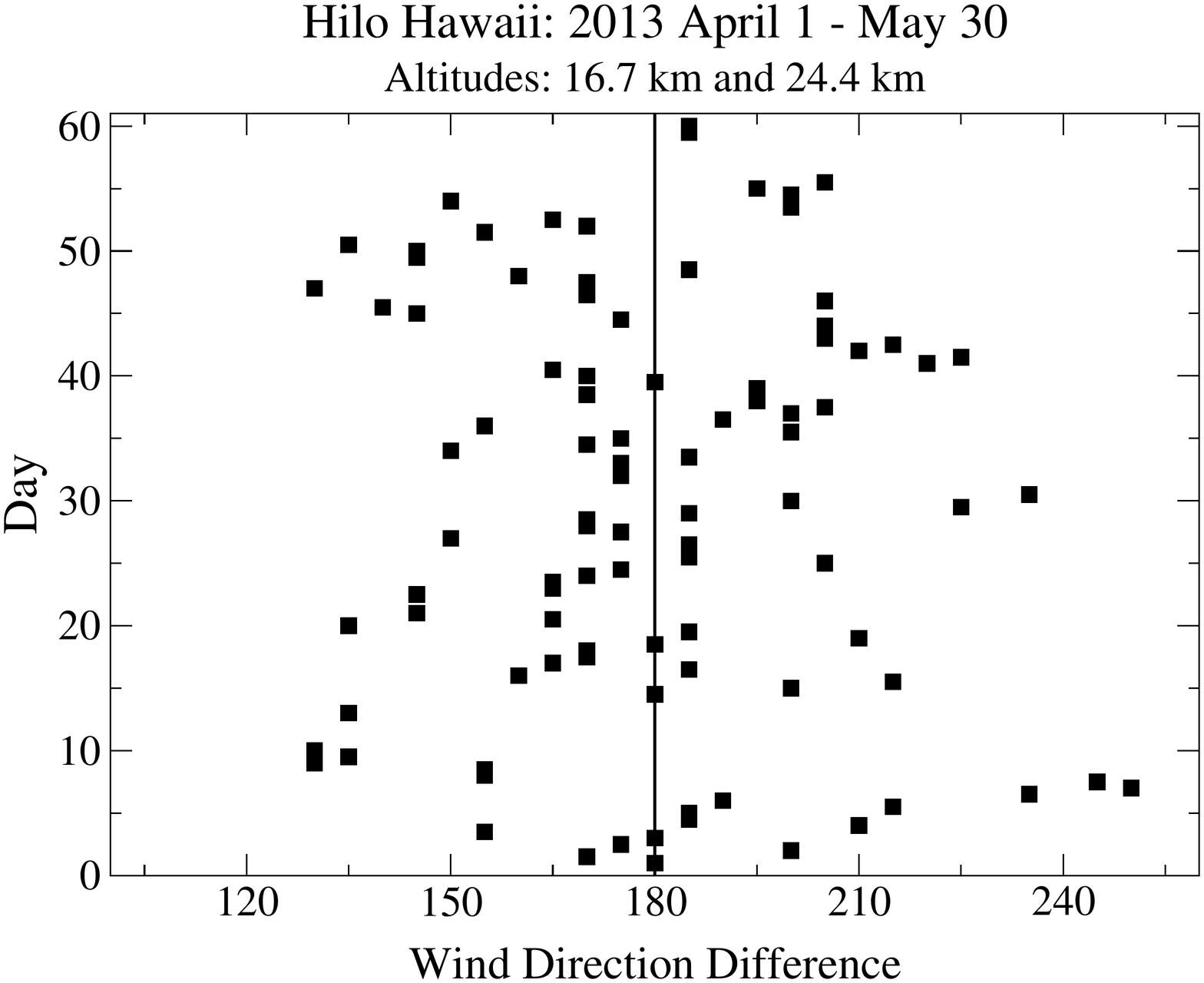}
\caption{Plots of Spring wind direction differences between 
stratospheric winds at 16.7 km and $24.4 \pm 1.0$ km above Hilo, Hawaii for four years 
based on radiosonde data. A difference of 180$^\circ$ (solid line) indicates directly opposing winds.  }
\label{fig:3}
\end{figure}

Although the plots of Figure \ref{fig:2} and the results of other stratospheric
wind studies \cite{Roberts11} indicate the existence of a usable stratospheric
wind shear, such multi-year average plots do not reflect the variable day-to-day wind
conditions that would actually govern the behavior of the proposed system.
Such day-by-day wind direction differences at altitudes of 16.7 and 24.4 km
(55 and 80 kft) are shown in Figure \ref{fig:3}.  Each of the four plots is for a
60-day interval in the spring of the years 2000, 2005, 2010, and 2013 for the
atmosphere above Hilo, Hawaii (latitude +19.8$^{\circ}$) and assembled from
radiosonde data available from the University of Wyoming's upper air sounding
website (http://weather.uwyo.edu).  Typically, two radiosonde flights are made
each day and both measurements are plotted when available.  The plot for 2013
shows the most recently available data.  More details may be found in
\cite{Roberts11}.

Data for a tug altitude of 16.7 km (55 kft) were extracted from the
radiosonde database within a relatively small altitude range ($\pm$ 0.2
km), while the airship's altitude was allowed to vary by $\pm 1$ km so as to reflect the
likelihood of altitude variations due to diurnal heating effects.  In cases of
missing radiosonde data within these altitude ranges, we interpolated between
the two closest values.  Although the data shown in Figure \ref{fig:3}  cover
an upper altitude range centered at 24.4 km, nearly 75\% of the measurements
plotted correspond to values taken at altitudes between 23.5 and 24.3 km.  

It is important to note that not all sounding data covering these time
intervals are shown in these plots.  Besides some missing radiosonde data (typically
just a few days during a month), we do not show wind direction differences
that exceed 70$^\circ$. Large variations in upper air flows can occasionally
lead to unfavorable wind conditions for several days each month. This is the
reason that during the year 2000 we show wind direction differences for March
16 -- May 15  rather than April 1 -- May 30. During that year, the wind
direction reversal formed over Hawaii about two weeks earlier than typically
seen.  During the four periods shown, the number of 12-hour periods during
which easterly and westerly wind direction were greater than 70$^\circ$ apart
were 23 in 2000, 18 in 2005, 15 in 2010, and 20 in 2013. However, on many of
these occasions, wind speeds were relatively low at one or both altitudes. 

Keeping in mind these limitations, the plots of Figure \ref{fig:3} illustrate
that between 16.7 and 24.4 km (55 kft and 80 kft) the stratospheric wind
directions are within 45$^\circ$ of being 180$^\circ$ apart for the majority of
the days shown. The best of these two-month periods is April and May 2005 when
over 85\% of the time the upper and lower altitude winds were within 30$^\circ$
of being 180$^\circ$ apart.  The worst two month period shown occurred in 2013.
Marked differences year to year is not surprising.  This is, after all,
weather, and weather patterns can change significantly from one year to the
next.  However, the regular appearance of such opposing wind flows between
stratospheric layers only some 7 km apart can be exploited to maintain the
geographical location of a high-altitude platform without the need of
propulsion power.

Because of seasonal wind variations above a particular geographic location,
stratospheric wind shear will not permit year-round station-keeping.  Suitable
opposing winds are found around $40^\circ$ latitude in hemispheric summers,
but in spring and fall they are found at lower latitudes around 15 to
25$^\circ$ (see Figure \ref{fig:2}.  This is shown in Figure 4 where we plot
wind direction differences at 15.2 km and 24.4 km (50 kft and 80 kft) for the
months of June and July in the years 2000 and 2010 over Denver, Colorado
(latitude +39.8$^{\circ}$).  Although there is considerable scatter, the lower to
upper stratospheric wind shear is still within 45 degrees of being directly opposite
over 75\% of the time. These plots also show that the wind shear can be experienced
by a tug at lower altitudes, here at 15.2 km (50 kft).

The seasonal shift in latitude of the stratospheric wind shear means that in
order to operate year-round the airship and tug  will need to move north or
south some 20 -- 30 degrees in latitude during the course of a year.  A shift
in latitude of the wind shear may be partially responsible for some of the
unfavorable wind shear days seen in the Spring months over Hawaii (see Fig.\
3).

\begin{figure}
\includegraphics[width=0.52\linewidth]{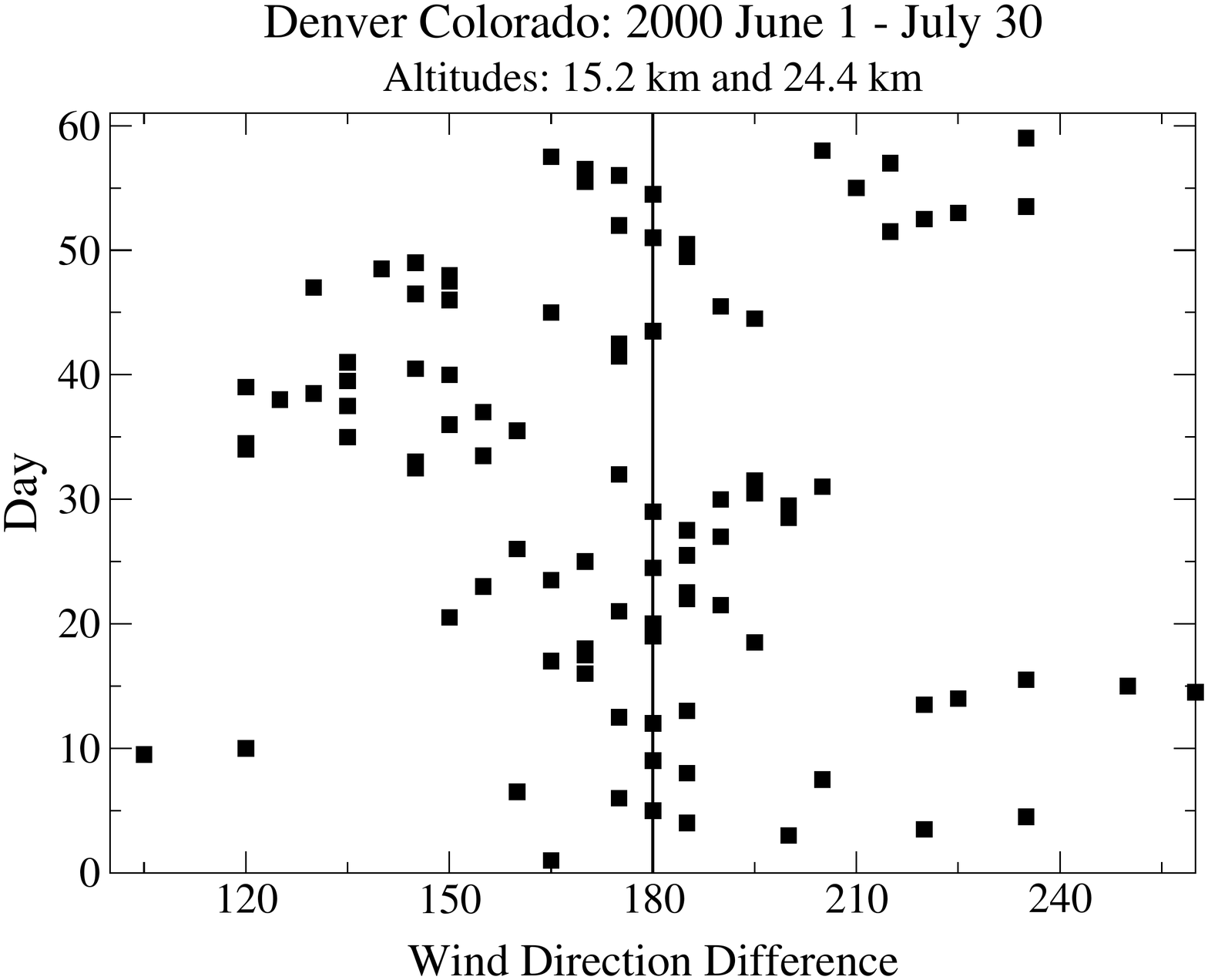}
\includegraphics[width=0.52\linewidth]{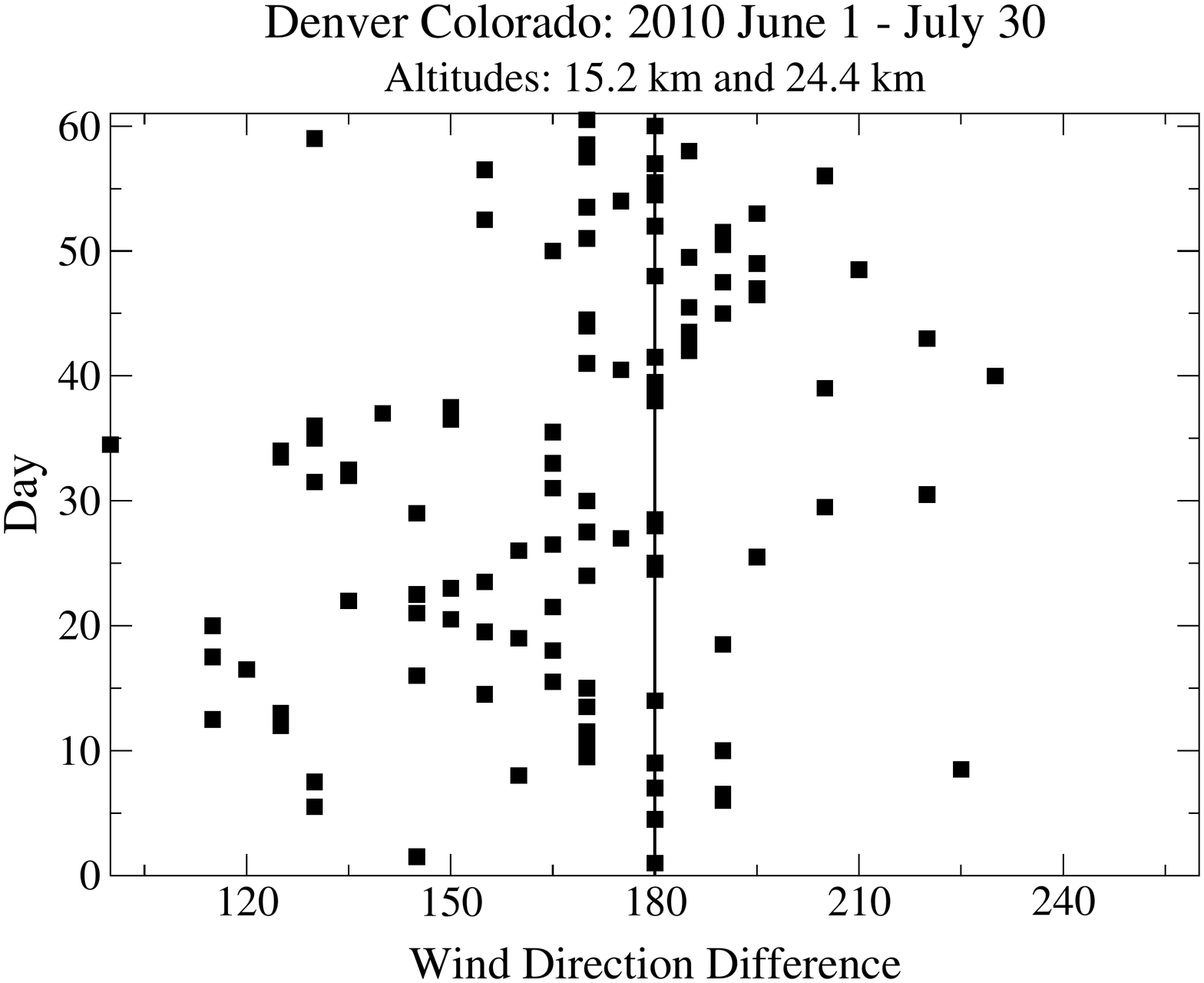} \\
\caption{Plots of Summer wind direction differences between
stratospheric winds at 15.2 km and $24.4$ km above Denver, Colorado for the years 2000 and 2010
based on radiosonde data. A difference of 180$^\circ$ (solid line) indicates directly opposing winds.  }
\label{fig:4}
\end{figure}

\section{Payload Platform and Tug Operation}

The operation of the proposed station-keeping system depends on balancing the
aerodynamic forces on the airship with those acting on the tug.  The air
density at the tug's altitude of 17 km is roughly three times that at the
airship's altitude of 24 km.  In addition, wind speeds are generally two to
four times greater at the lower altitude than at the higher altitude.  The tug
will therefore experience drag forces some 10 to 50 times higher than the
airship even if the two vehicles are of similar size and shape.  If the airship
is streamlined so as to minimize drag, the tug could be made quite compact and
lightweight while still developing the counter force necessary for hold the
airship steady in the wind.

Table \ref{shipforce} shows sample wind induced drag force calculations for
three airship sizes and shapes.  These computations assumed an airship altitude of 23.7 km
(78 kft) and a variety of ambient wind speeds.  
The listed drag force values were
calculated assuming only form and surface drag.  Case 1 is an airship  similar
in size to SwRI's streamlined HiSentinel80 airship, Case 2 is a ``super-sized''
HiSentinel80, and Case 3 is for a spherical balloon having a displaced volume 
similar to that of HiSentinel80 in Case 1.

Comparing Cases 1 and 3, it is clear that having a streamlined airship versus
a spherical balloon lowers the total wind drag force by about a factor of
20.  Also, going from a small to a larger streamlined airship (Cases 1 and 2)
the system gains a factor of nearly 10 in potential lift while the total drag
force increases by only a factor around 2.5.

\begin{table}
\caption{Airship Drag Forces at 23.7 km (78 kft), $\rho = 0.048$ kg/m$^{3}$ }
\label{shipforce}       

\textbf{Case 1:} HiSentinel80: D = 15 m, L = 60 m; volume: 10600 m$^{3}$  \\
balloon: 320 kg (@ 0.1 kg/m$^{2}$); helium: 70 kg; displaced air: 510 kg \\
\begin{tabular}{ccccc}
\noalign{\smallskip}
\hline\noalign{\smallskip}
  Drag    &     &   & Wind Speed  & \\
 Parameters &  ~ ~ 5 m/s ~ & ~ ~ 10 m/s ~ &  20 m/s ~ & ~ 30 m/s ~  \\
\noalign{\smallskip}
\hline\noalign{\smallskip}
C$_{D}$ = 0.03; A$_{f}$ = 177 m$^{2}$      &    3 N   &  13 N &   ~ 51 N &  114 N \\
C$_{SF}$ = 0.003; A$_{w}$ = 3500 m$^{2}$  &    6 N   &  25 N &    100 N &  227 N \\
\noalign{\smallskip}
\noalign{\smallskip}
Total Drag Force  &  9 N   &  38 N &   151 N &  341 N \\
\noalign{\smallskip}\hline
\noalign{\smallskip}\hline
\noalign{\smallskip}
\end{tabular} \\
\textbf{Case 2:} Super-HiSentinel: D = 25 m, L = 100 m; volume: 49100 m$^{3}$ \\ 
balloon: 885 kg (@ 0.1 kg/m$^{2}$); helium: 325 kg; displaced air: 2350 kg \\
\begin{tabular}{ccccc}
\hline\noalign{\smallskip}
  Drag    &     &   & Wind Speed  & \\
 Parameters &  ~ ~ 5 m/s ~ & ~ ~ 10 m/s ~ &  20 m/s ~ & ~ 30 m/s ~  \\
\noalign{\smallskip}\hline\noalign{\smallskip}
C$_{D}$ = 0.03; A$_{f}$ = 490 m$^{2}$      & ~ ~9 N   & ~ 35 N   & ~141 N & ~ 317 N \\
C$_{SF}$ = 0.003; A$_{w}$ = 9000 m$^{2}$  & ~ 16 N   & ~ 65 N  & ~ 260 N & ~ 583 N \\
\noalign{\smallskip}
\noalign{\smallskip}
Total Drag Force  & ~ 25 N   & ~ 100 N  & ~ 401 N & ~ 900 N \\
\noalign{\smallskip}\hline
\noalign{\smallskip}\hline 
\noalign{\smallskip} 
\end{tabular} \\
\textbf{Case 3:} Spherical Balloon: D = 28 m; volume: 11500 m$^{3}$ \\
balloon: 250 kg (@ 0.1 kg/m$^{2}$); helium: 75 kg; displaced air: 550 kg \\
\begin{tabular}{ccccc}
\hline\noalign{\smallskip}
  Drag    &     &   & Wind Speed  & \\
 Parameters &  ~ ~ 5 m/s ~ & ~ ~ 10 m/s ~ &  20 m/s ~ & ~ 30 m/s ~  \\
\noalign{\smallskip}\hline\noalign{\smallskip}
C$_{D}$ = 0.5; A$_{f}$ = 615 m$^{2}$      & ~ ~185 N   & ~ 740 N   & 3000 N & ~ 6700 N \\
C$_{SF}$ = 0.003; A$_{w}$ = 2460 m$^{2}$  & ~~ ~4 N   & ~ 18 N  & ~ 70 N & ~ 160 N \\
\noalign{\smallskip}
\noalign{\smallskip}
Total Drag Force  & ~ 190 N   & ~ 760 N  & ~ 3100 N & ~ 6900 N \\
\noalign{\smallskip}\hline
\hline\noalign{\smallskip}
\end{tabular}
\end{table}

The drag values listed in Table \ref{shipforce} must be comparable to the wind
drag numbers for the lower altitude tug vehicle which are listed in Table
\ref{tugforce} for a range of wind speeds likely to be encountered at the tug's
altitude of around 17 km.  As an example, we have adopted a tug design in
the form of a conventional glider consisting of a narrow fuselage and thin,
high-aspect wings with high lift-to-drag ratios. We have assumed
some sort of variable drag device as part of the tug with a form drag force proportional to an
adjustable area of the device.  The table shows drag force values resulting
from both open and closed configurations.

As Table \ref{tugforce} shows, it appears feasible for a tug to generate drag
forces covering the complete wind speed range calculated for the two
streamlined airship cases in Table \ref{shipforce} (Cases 1 and 2) but not for
a spherically shaped airship (Case 3).  This again disfavors a spherical
airship shape.

\begin{table}
\caption{Tug Vehicle Drag Forces}
\label{tugforce}       
Tug: D = 0.75 m, L = 4 m fuselage + 4 m drag device, altitude = 16.7 km (55 kft) \\
\begin{tabular}{ccccc}
\noalign{\smallskip}\hline
\hline\noalign{\smallskip}
Vehicle      &  Drag      &          & Wind Speed    &          \\
Component    & Parameters &  ~ 10 m/s & ~ 20 m/s     &   ~ 30 m/s  \\
\noalign{\smallskip}\hline\noalign{\smallskip}
fuselage + wings           & C$_{D}$  = 0.12;  A$_{f}$ =  1.8 m$^{2}$  & ~ 2.0 N  & ~ 9.0 N & ~ 20 N \\
                           & C$_{SF}$ = 0.03;  A$_{SF}$ = 7.1 m$^{2}$  & ~ 2.0 N  & ~ 8.0 N & ~ 19 N \\
drag device closed  & C$_{SF}$ = 0.03;  A$_{SF}$ =  7.1 m$^{2}$  & ~ 2.0 N  & ~ 8.0 N & ~ 19 N \\
~ " ~ ~  "  ~ opened  & C$_{D}$  = 1.0;   A$_{f}$ =  28 m$^{2}$      & ~ 270 N  & ~ 1090 N & ~ 2450 N \\
\noalign{\smallskip}\hline
\noalign{\smallskip}
Total Force Range     &                                           & 6 - 275 N & 26 - 1115 N & 60 - 2490 N \\
\noalign{\smallskip}\hline
\hline\noalign{\smallskip}
\end{tabular}
\end{table}

\begin{table}
\caption{Tether Masses and Wind Loads at 20 m/s  }
\label{tetherforce}       
Single Tether: Altitude: 0 to 20 km; Total Tether Length: 20 km; C$_{D}$  = 1.0  \\
\begin{tabular}{crrrrrr}
\noalign{\smallskip}\hline
             &               &               & Altitude       &            &                         \\
   Dyneema             & 0 - 5 km    & 5 - 10 km      &  10 - 15 km   & 15 - 20 km   &   \\
    (SK78)          & 0.96 kg/m$^{3}$ & 0.56 kg/m$^{3}$ & 0.30 kg/m$^{3}$ & 0.13 kg/m$^{3}$ &   Totals   \\
\noalign{\smallskip}
\hline\noalign{\smallskip}
  5mm; MBS 3300 kg  &  480 kg     &  280 kg       &         &           &    760 kg           \\
  mass: 15 kg/km    &   75 kg     &   75 kg       &         &           &    150 kg           \\
 3mm: MBS 1400 kg   &             &               &   90 kg &     40 kg &    130 kg           \\
  mass: 5 kg/km     &             &               &   25 kg &     25 kg &  \underline{~~~50 kg}            \\
                     &              &               &               &            &     1090 kg                 \\
\noalign{\smallskip}\hline
\hline
\smallskip
\end{tabular}
Tug - Airship: Altitudes: 17 and 24 km; Total Tether Length: 12 km; C$_{D}$  = 1.0 \\
\begin{tabular}{crrrrrr}
\noalign{\smallskip}\hline
             &   ~~~~~~~~~~            &               & Altitude        &            &                         \\
   Dyneema          &      ...    & ~~~~~~~~~~~...      & ~~~~~~ 17 - 21 km      & 21 - 24 km    &     \\
    (SK78)          &      ...    & ~~~~~~~~~~~...      & ~~~~~~0.10 kg/m$^{3}$ & 0.06 kg/m$^{3}$ &   Totals   \\
\noalign{\smallskip}
\hline\noalign{\smallskip}
\noalign{\smallskip}
 2mm: MBS 450 kg     &  ~~~~~     & ~~~~~   &   ~~~~ ~   16 kg  &  ~~~~ 7 kg &~~~    23 kg           \\
  mass: 2.4 kg/km    &            &         &            17 kg  &      12 kg   &  \underline{~~~~~29 kg}            \\
                     &            &         &                   &             &      52 kg             \\
\noalign{\smallskip}\hline
\end{tabular}
\end{table}

Lastly, we show in Table \ref{tetherforce} estimated wind loading values for
both a ground-tethered high altitude aerostat and our proposed high altitude airship-tug
scheme. Here we have assumed a constant wind speed of 20 m/s at all altitudes.
Although there are a variety of tether materials that could be used in either
scheme, for these sample calculations we chose Dyneema SK78 as the tether
material. There are stronger tether material options
which have higher breaking strengths but these numbers serve to give a sense
of mass and wind loads at various altitudes and hence required tether strength.

For the single ground-tethered scheme, we chose a 5 mm tether for altitudes 0 --
10 km (0 - 33 kft) and a 3 mm tether for altitudes 10 -- 20 km (43 - 65 kft). 
A thicker tether might be required at
lower altitudes since it transverses the whole troposphere where more 
severe transient wind loads are likely to be experienced.  In contrast, a thinner 2
mm cord was chosen for the the airship-tug tether since wind loading conditions
are far more benign above the jet stream and most storms at an altitudes 17 km (55 kft)
and higher.

Comparison of the two high-altitude tethered airship approaches in Table
\ref{tetherforce} shows that a single tether will experience just under one
metric ton of horizontal wind loading plus tension due to 200 kg of tether
mass. Although this estimate assumes a constant wind speed of 20 m/s along the
entire 20 km length, these wind loads and tether mass could actually be an
underestimate.  It is unlikely that the tether would be as short as 20 km given
wind loading and varying wind directions and speeds from the ground winch up to
the altitude of 20 km, and thus a tether length as much as 30 km is probably
more realistic. In that case, again dividing the tether into 3 mm and 5 mm
thicknesses---but now each 15 km long---an even greater wind loading might
exist while the total tether mass increases to around 300 kg. 

In real life, the situation might be even less favorable since tropospheric
wind speeds often exceed 20 m/s and can even be over 50 m/s in the jet stream.
At a wind speed of 40 m/s, just a  1 km long section of a 5 mm thick tether at an
altitude around 10 km (30 kft; $\rho = 0.4$ kg/m$^{3}$) would have a wind load
of 150 kg for this short section. 

In any case, a minimum break strength (MBS) safety factor for a single tether
with the chosen thicknesses is low and much less than the usually desired
factor of 5 or more. Thus, such tether weight and wind loading estimates would
seem to pose serious operational challenges for maintaining a stratospheric
airship with a grounded tether complicating the ground tether approach further.

While, as in the single tether case, a considerably longer tether will be
needed in reality than just the 7 km altitude separation of airship and tug, a
shorter and thinner tether in an airship-tug scheme offers both a lower tether
mass and wind loading.  A tether of SK78 Dyneema 12 km long will have a
combined mass load and wind load well below the tether's MBS of 450 kg. For
example, even a relatively high 30 m/s wind speed over the entire 12 km long 2
mm tether at altitudes between 17 and 24 km will only generate a total wind
load of less than 100 kg.  

\section{Discussion}

The airship-tug station-keeping arrangement discussed above uses the naturally
occurring seasonal stratospheric wind shear to provide the needed energy to
keep the system on station.  The payload carrying platform's altitude around 80
kft is also much higher than that of a self-propelled airship at 65 kft thereby
providing wider horizon to horizon coverage of the Earth and better upward
viewing image quality.  This tether scheme also avoids several problems
associated with a ground-based tethered platform; namely, little if any
aviation hazard, no winch, no stormy weather to fly through, and a shorter
tether meaning less tether weight and wind loading.  In addition, the tether is
expected to be always under some tension so slack issues that can arise in a
ground-based winch tether arrangement are reduced.  Wind loading at altitudes
above 15 to 17 km (50 to 55 kft) should also be relatively low even in high
wind conditions, making a thin and lightweight tether practical.

There are several key components of the concept that will determine its
reliability and effectiveness.  The higher-altitude LTA platform must be
constructed so as to have no appreciable fabric or seam leaks of lifting gas
(i.e., hydrogen or helium) thus permitting long float durations of
weeks to months. Both it, the tug and the tether 
should be as lightweight as possible enabling the greatest payload mass in
relation to the balloon's lift capability.

Ideally, the upper LTA platform would also have a streamlined aerodynamic shape
so as to lessen wind drag forces as much as possible.  It should also have some
directional lift capability such as through a rear vertical stabilizer so as to
help steer it into or against the prevailing winds and be designed for
flexibility in payload mounting configuration.  For example, astronomers may
want a top-mounted telescope that has unobstructed access to targets near the
zenith, while Earth scientists may prefer down pointing instruments.

However, the most critical component of the proposed concept is perhaps the tug
vehicle.  We conceive the tug as taking the form of a ultra-lightweight glider
with intrinsically low drag.  It could develop the forces needed to counter
drift of the airship in two ways: deployment of a variable drag device such as
a parachute or umbrella like device or variable pitch propeller(s), or it could
generate appropriate aerodynamic forces with its wings.  

Drag is necessarily in the direction of airflow, so it may seem that the
high-drag configuration would only work if the winds at the two altitudes
exactly oppose. But if the airship were a ``dirigible" design, it could develop
aerodynamic forces that are not precisely parallel to wind direction.
Similarly, the tug could be controlled to fly in a direction that produced the
necessary tether force over a wide range of angles relative to the wind
direction.  

The combination of a semi-steerable LTA airship and a maneuverable drone-like 
tug with variable lift capability could allow the system to keep
station in a variety of wind combinations.  It could even maneuver to find
better wind conditions, and climb and descend to some degree as needed.

The tug will need to be able to generate it own power to serve its operating
flight systems and possibly to be self-propelled to some extent.  In addition
to solar PV power stored in batteries, the tug could be equipped with a propeller
to serve as a variable drag device and power from the propeller could be used to
generate electricity both day and night.  

A cruder wind force balancing scheme was proposed in 1969 by R.
Bourke \cite{Bourke69} in a Raytheon Company report. He described a concept in
which a conventional balloon floating in the stratospheric easterlies could deploy a 
parachute into the lower stratospheric westerlies to provide a drag force to overcome 
the balloon's drift.

Using available wind data available at the time, Bourke concluded that this
arrangement could work for certain months of the year, mainly during summer
months at mid-latitudes. But he also found that the altitude and latitude of
the lowest stratospheric winds varied seasonally leading to difficulties in
maintaining accurate station keeping. Nonetheless, he viewed the concept as
``provocative in its intrinsic simplicity''.  However, to our knowledge no
high-altitude balloon plus drag chute system was ever deployed and tested by
Raytheon or anyone else.

Our scheme differs substantially from that proposed by Bourke.  He suggested
that the upper altitude balloon have self-propulsion capabilities and proposed
a simple drag chute lowered from the balloon using a winch only as a
supplemental element to aid the airship's station-keeping ability. In contrast,
our concept consists of a passive and ideally aerodynamically-shaped,
stratospheric balloon or airship tethered to a lower altitude robotic tug
vehicle that can precisely control its aerodynamic wind forces.  Our
stratospheric airship would have no self-propulsion element but could have some
directional steering capabilities much like that demonstrated in a high
altitude wing guidance system \cite{Nock07}.  Bourke's use of a winch-lowered
drag chute may have been an attempt to simplify the balloon launch.  Our scheme
could also include some sort of tether storage system possibly attached to the
tug vehicle in an effort to better control deployment and recovery of both
upper and lower vehicles.  

\section{Astronomical Uses of a High Altitude Platform}

One application for a stratospheric platform would be wide-field, high
resolution optical and near-infrared imaging of astronomical targets.  The
value of high angular resolution imaging for astronomy cannot be overstated.
The chief reason for the enormous impact of the Hubble Space Telescope (HST)
across a wide spectrum of research topics despite its modest size mirror (2.4
m) has been its ability to obtain diffraction-limited imaging due to its
location above Earth's atmosphere.  

However, with no repair or refurbishment missions currently planned, Hubble's
expected useful lifetime will probably end before the year 2020 due instrument
failures or degradation of its batteries, solar panels, pointing gyros, and
associated equipment \cite{Moskowitz}.  With no present follow-up optical/UV
space mission to Hubble, its loss may mean astronomical high-resolution imaging
might be confined for the near future to small space telescopes or ground-based
adaptive optics (AO) instruments which employ one or more natural or laser
guide stars to correct for atmospheric turbulence.  Unfortunately, AO
instruments work best in the infrared and under good seeing conditions and
provide limited field-of-view ($<$ 1 arcmin) with Strehl ratios less than 60\%
\cite{Rigaut14}.

A reliable LTA platform situated at an altitude of 20 km or higher should, if
properly equipped, provide image quality competitive with space-based
telescopes.  Such an observatory could provide sub-arcsecond imaging with short
response times at a much lower cost than a comparable space-based telescope.  

For example, at an altitude of 24 km (80 kft) an astronomical telescope would
be above the weather and all but $\approx$ 2.5\% of the atmosphere. It would experience
virtually perfectly clear skies every night with image quality at or
approaching the diffraction limit of the main aperture. Thus, an optical
telescope located at such stratospheric altitudes with a mirror just 0.5 m in
diameter (20-inch) with sufficient pointing stability and large CCD arrays
could provide wide-field images with FWHM = 0.25 arcsecond at 500 nm, making it
virtually superior to any ground-based imaging system.

Being above the weather, it could provide such data quality night
after night for as long as the platform remained at this altitude.  The lack of
appreciable water vapor, dust and other particulates in the remaining
atmosphere above these altitudes such a platform would also enjoy excellent
atmospheric transmission.

Light scattering from moonlight would be expected to be minimal and not a major
factor in scheduling faint target observations, making most observing time
effectively astronomical ``dark time.''  This feature would greatly enhance the
platform's ability to respond rapidly to opportunities for observations of faint
transient targets such as supernovae and gamma-ray bursters.  

Also, unlike low Earth orbit (LEO)  satellites such as HST, data transfer to and from a
high-altitude station-keeping observatory could involve simple line-of-sight
communications running continuously to a single ground station.  Finally,  a
stratospheric astronomical observatory could also provide reliable science
support for a host of space-based missions at an estimated cost of a few
percent of a conventional LEO satellite.

\section{Conclusions}

We have described a new method for establishing a near station keeping,
stratospheric LTA vehicle at low and mid-latitudes.  This concept uses the
naturally occurring seasonal wind shear between upper and lower layers of the
stratosphere to provide forces that counter platform wind drift and allow it to
keep station over a specified geographical location.  We have necessarily
left out many details about the architecture. These include platform
migration issues in order to follow seasonal variations in latitude where
optimal stratospheric wind shears are found, launch and recovery problems and
solutions, specific airship and tug design constraints, and science payload
arrangements to permit unobscured horizon-to-horizon observations.

If this method is shown to be practical, then the quest for the long-sought
method of station keeping a scientific HALE platform may finally be realized,
within season and latitude restrictions.  This concept could provide the means
for obtaining high quality data rivaling space-based platforms but at a small
fraction of the cost. The development of an affordable stratospheric platform
that could keep station for weeks or months would be a powerful new tool for a
variety of users and could be a game-changer for astronomical, atmospheric, and
Earth-science research, as well as for a host of other applications including
military surveillance and civil telecommunications services.

\begin{acknowledgements}

The authors gratefully acknowledge valuable advice and conversations
about high-altitude LTA science platforms from participants in the W. M.
Keck Institute for Space Studies (KISS) workshop entitled ``Airships: A New
Horizon for Science,'' especially Jeff Hall, Steve Lord, Steve Smith, Mike Smith, and
workshop co-leads Sarah Miller, Lynne Hillibrand, and Jason Rhodes.

\end{acknowledgements}


{}

\end{document}